\documentstyle[preprint,revtex]{aps}
\tightenlines
\def\div{{\rm div}}
\def\rot{{\rm rot}}
\def\v{{\bf v}}
\def\r{{\bf r}}
\def\w{\vec{\omega}}
\def\i{{\bf i}}
\def\j{{\bf j}}
\def\k{{\bf k}}
\def\ux{\hat{\bf x}}
\def\uy{\hat{\bf y}}
\def\uz{\hat{\bf z}}
\def\half{\frac{1}{2}}

\def\f{{\bf f}}
\def\x{{\bf x}}
\def\y{{\bf y}}
\def\z{{\bf z}}
\def\dx{\delta x}
\def\dy{\delta y}
\def\dz{\delta z}
\def\di{\delta i}
\def\dj{\delta j}
\def\dk{\delta k}
\def\px{\frac{\partial}{\partial x}}
\def\py{\frac{\partial}{\partial y}}
\def\pz{\frac{\partial}{\partial z}}
\def\pii{\frac{\partial}{\partial i}}
\def\pj{\frac{\partial}{\partial j}}
\def\pk{\frac{\partial}{\partial k}}
\def\b{{\bf b}}
\def\p{{\bf p}}
\def\g{{\bf g}}
\def\q{{\bf q}}
\def\v{{\bf v}}

\def\vx{v_{x}}

\def\wx{\omega_{x}}

\def\vxx{\partial \vx / \partial x}
\def\vxy{\partial \vx / \partial y}
\def\r{{\bf r}}
\def\x{\hat{\bf x}}
\def\y{\hat{\bf y}}
\def\z{\hat{\bf z}}
\def\i{\hat{\bf i}}
\def\j{\hat{\bf j}}
\def\k{\hat{\bf k}}

\begin{document}
\draft
\preprint{ Turbulence simulated with the lattice model II}
\begin{title}
Homogeneous Isotropic Fluid Turbulence\\
Simulated with the Lattice Vortex Tube Model\
\end{title}
\author{Y-h. Taguchi}
\begin{instit}
 Institut f\"ur Festk\"orperforschung,
 Forschungszentrum J\"ulich,
 D-5170 J\"ulich, Germany;\\
 Department of Physics,
Tokyo Institute of Technology,
Oh-okayama, Meguro-ku,\\
 Tokyo 152, Japan\cite{present}
\end{instit}
\author{Hideki Takayasu}
\begin{instit}
 Department of Earth Science, Kobe University, Kobe 657, Japan.
\end{instit}

\begin{abstract}
Fully developed turbulence is analised with the lattice model
employing vortex tube representation which
is introduced recently by the authors.
Several characteric features observed in experiments and direct numeric
integrations are reproduced. Not only Kolmogorov's inertial range is
observed, but also several local probability
distribution functions  are obtained as well.
Those of the local velocities are close to the Gaussian and
exponential-like distributions appear in local vorticity,
relative velocities and local velocity consisting of only higher
wave number components.
Coherent structure of vortex tubes is seen, too.
Moreover required cpu-time and memory-size are very little comparing
with the conventional pseudo-spectral method.

keywords : fluid turbulence, vortex tube, lattice model, numerical technic.
\end{abstract}

\section{Introduction}

Fully-developed turbulence is one of the important topics in fluid dynamics.
Although it has been studied over a century,
we have not yet gotten any clear concept of what turbulence is.
Even if we consider only the simplest case of 'Homogeneous Isotropic
Turbulence',
we do not  know what the essential point is.
This is mainly because there has been no powerful tool
 to investigate it on either experiment or theory.
In the experiments, the data including most information is no more than
flow lines of trace-particles on the two dimensional sub-section,
 which is far from enough to investigate the fluid in details.
Theorists have no general method to  deal highly nonlinear systems with
many degrees of freedom which the turbulence belongs to.
Even using numerical methods, it is very hard to get the solution of
fully developed turbulence because of 'fractal property' of
it\cite{Paladin}.
If a system has fractal nature, we cannot find any characteristic length.
In such an occasion, we must calculate on a large system to include
many kinds of scale length.
This requires us to use large lattice sizes (or many number of wave numbers)
which needs a long cpu times on computer.
Even the largest and fastest computers cannot give us
the numerical solution of fully developed turbulence,
although recently some part of characteristic features
can be reproducible on computer\cite{Kida89,Metais,Hosokawa88,Hosokawa89}

In this paper, we propose to give a lattice model  which can reproduce
some of characteristic feature of statistical property of turbulence, ex.
scaling properties etc.
If we succeed in constructing a simple model showing some behaviors which are
 similar to the fluid,
we can say that the origin of them is in the common parts.
Therefore it will  help us to know what the essence of turbulence is.

The paper will be organized as followings.
In Sec.~\ref{model : sec} we will denote on which view point we stand and
the description of fluid on 'lattice' will be defined.
In Sec.~\ref{dyna : sec}, we will define a dynamics (time-marching) of our
model
and show it is equivalent to the Navier-Stokes equation.
The results and discussion will be presented in
Sec.~\ref{results : sec} and Sec.~\ref{disc : sec} will
provide us a summary and a conclusion.

\section{Modeling fluid on the lattice}
\label{model : sec}
\subsection{Concept}

The concept which we employ in this paper is a vortex tube representation of
fluid turbulence.
This is because it can easily provide us a lattice model to simulate
turbulence.

Recently, in the field of statistical physics,
simple lattice models have been shown to be good tools
to reproduce qualitative features of non-linear systems
even if they were essentially continuous ones.
For example, the diffusion limited aggregation on lattices
has turned out to be a good model to reproduce
 fractal dimension of the Laplacian growth systems which
 are essentially continuous \cite{Meakin}.
K. Kaneko\cite{Kaneko} has proposed coupled map lattices
in order to study chaos in the many degrees of freedom systems.
And Y. Oono\cite{Oono} has proposed the cell dynamical system on lattices and
succeeded in getting good results on the phase separation process,
which has been believed to be described by the Time-Dependent-
Gintzburg-Landau eq.; a sort of partial differential equation.
And one of the author (H.T.) has also proposed many
discretized models\cite{Matsuzaki,Takayasu85a,Takayasu85b} and
obtained results consistent with the experimental data.
These lattice models are more suitable to be treated on computer
 than conventional method like partial differential equations.
Following this line we purpose to construct a lattice model which can
reproduce qualitative features of turbulence in this paper.
To construct these lattice models, we need some physical quantities
whose local dynamics are well known.
And introducing  interactions between them,
we can get a complete system to represent a real system.
For this purpose, the vortex tube representation is very convenient.
The vortex lines are believed to move with the fluid
because of the conservation law of circulation.
In the following, we will construct the lattice model so as to
preserve these essential features.
The first step of this process is to answer the question;
'How can we map the fluid on the lattice?'

\subsection{The representation of the fluid on the lattice}

To map the fluid on the lattice,
we must take special care for
some basic values on the vector analysis.
For example, the divergence of velocity $\v$,
 $\div \v$, must be easy to calculate,
because incompressibility condition requires $\div \v =0$.
And  $\rot \v$ also should be calculated easily because of
the definition of the vorticity $\w$;$\rot \v = \w$.
Although there is no uniqueness on how to map the fluid on the lattice
to satisfy these requirements,
we employ the following description.

First, we map the velocity field $\v(\r)$ on each bond
on the simple cubic lattice (we will call it as the $\v$-lattice),
such that $x$ bonds must have the only $x$ component of the velocity.
$y$ and $z$ bonds also should have the only component along the bond direction.
On the other hand, the vorticity field $\w(\r)$ is assigned on
each bond on the dual lattice(we call it $\w$-lattice).
Again each bond has only one component.
This means, on the $\v$-lattice, the vorticity is located on the
center of each face and its direction is perpendicular to the face
(See Fig.~\ref{lattice}).
For example, the position vector of the bond between the lattice points
$(x_0, y_0, z_0)$ and $(x_0+1, y_0, z_0)$ is written as
$(x_0+1/2, y_0, z_0)$.
In this description, the vorticity field $\w(\r)$ is
related to the velocity $\v(\r)$ by the equation

\begin{equation}
\omega_i(\r) = v_k(\r + \half \j)+ v_j(\r - \half \k)
- v_k(\r - \half \j) -  v_j(\r+\half \k), \label{rotv},
\end{equation}

where $\i,\j,\k$ are unit lattice vector of $x,y,z$
directions, which can be exchanged their order cyclically.
Therefore $\w$ is circulation rather than the vorticity.
However we are not interested in the difference between them,
because they share the same basic property mentioned in the
previous subsection, for example folding etc.
Independent of whether our $\w(\r)$ is the vorticity or the circulation,
we get the same description.

Now, we have everything necessary to construct the map to the lattice model.
For example, $\div \v(\r)$ is given as

\begin{equation}
\div \v(\r) = \sum_{\i=\ux,\uy,\uz} \pm \v(\r\pm\half \i)\label{divv},
\end{equation}

where $\ux,\uy$ and $\uz$ are the unit lattice vectors of $x,y$ and
$z$ direction respectively.
However in this description, we have a problem.
If we give $\v(\r)$ first, then we can calculate $\w(\r)$
without any problems.
But if $\w(\r)$ is given first, we can not obtain $\v(\r)$ uniquely.
As will be shown, this is necessary for
time evolution of our model.
In addition to this, we xrequire the incompressibility
$\div \v(\r)=0$.
To get $\v(\r)$ satisfying this condition seems not to be trivial.

The basic procedure to calculate $\v(\r)$ from $\w(\r)$ is
essentially solving a set of linear equations given by (~\ref{rotv}~).
One may think that it is a similar task to solving the Laplace equation.
However, this is much more difficult because we cannot use the
relaxation method.
The method which we employ is the cg method which is
explained briefly in Appendix~\ref{ap1 : sec}.
In principle, this method can give us accurate numerical solutions
even for large matrices.
However, if we assume the periodic boundary condition
for the lattice, we have another problem.
Because of the periodicity, one of a set of linear equations mentioned above
is not independent to the rest equations.
Although we cannot prove this point analytically,
the rank of coefficient matrix of the set of linear equations
is always less than the number of components by unity.
(The total number of the components is equivalent to
the total number of independent components on the $\v$ lattice.
This is $3\times L^{3}$ if the linear dimension of the lattice is $L$.)
Therefore we have one degree of freedom on ambiguity of values of $\v(\r)$.
This ambiguity is used to solve the
last problem ; to fix $\v(\r)$ so as to
satisfy $\div \v(\r)=0$.
It looks a difficult task because we have only one degree of
ambiguity of the velocity field.
Fortunately this problem turns out to be solved by requiring a
 discretized representation of $\div \v = 0$
at an arbitrary point $\r_0$.
 This gives us the coefficient matrix whose dimension is equal to its rank.
 The velocity field $\v(\r)$ which this matrix gives
 turns out to satisfy
$\div \v=0$ and $\rot \v =\w$.
 Therefore we succeed in mapping
  fluid on the lattice.

\section{Dynamics of the lattice model}
\label{dyna : sec}
In the previous section, we have finished introducing basic elements
 to the lattice model; $\w(\r)$ and its 'interaction' $\v(\r)$.
 Actually, our model corresponds to a set of the vortex tubes moving
 with the velocity induced by the Biot-Savier interaction.
 In this section, we define a concrete process of dynamics.

In our model, vortex tubes are represented as a sequence of
bonds having non-zero vorticity.
It can be folded and have branches but no ends.
This character must be held even while the vortex tubes are transferred
by the flow of the fluid.
The simplest dynamics to satisfy these condition is
to move the vorticity on some $i$-bond to the neighboring $i$-bond along
$j$-direction
by the amount of $\omega_i(\r)v_j(\r)\Delta t$ and to
add some vorticity to neighboring $j$-bonds in order to have no ends.
However in stead of that, to move half of the net flow forward and the rest
half
backwards is more favorable in order to recover isotropy
(See the Appendix~\ref{ap2 : sec}).
If we consider $i=j$ makes no change, the time development is defined as;
\begin{eqnarray}
\omega_z(\r_0,t+\Delta t )  & = & \omega_z (\r_0,t)+\Delta t
[\pm J_{yz}(\r_0 \mp \hat {\y}, t) \pm J_{xz}(\r_0 \mp \hat {\x}, t)\nonumber\\
& &  \mp J_{zx}(\r_0-\hat {\z} / 2  \mp \hat {\x} / 2 , t)
 \mp J_{zy}(\r_0-\hat {\z} / 2  \mp \hat {\y} / 2 , t)\nonumber\\
& &  \mp J_{zx}(\r_0 + \hat {\z} / 2  \mp \hat {\x} / 2 , t)
 \mp J_{zy}(\r_0+\hat {\z} / 2  \mp \hat {\y} / 2 , t) ]/2 \nonumber\\
\label{dyna},
\end{eqnarray}
where $J_{ij}(\r,t)= \tilde {v}_i(\r,t)\omega_j(\r,t)$
and $\tilde{v}_i(\r)$ is the $i$-th component of the velocity
on the $\w$-lattice and is defined
as the average over neighboring bonds in the $\v$-lattice.
For example, $\tilde{\v}$ on $z$-bond is defined as
\begin{displaymath}
\tilde{ \v }( \r_0 )=([v_x(\r_0+\hat {\y}/2)+v_x(\r_0-\hat {\y}/2)]/2,
[v_y(\r_0+\hat {\x}/2)+v_y(\r_0-\hat {\x}/2)]/2, 0)
\end{displaymath}
$y$ and $z$ components have similar time developments.
$\Delta t$ is the time interval and chosen so as to $\max_\r
\mid J_{ij}(\r,t) \Delta t\mid = \alpha$.
This makes the change of the strength of $\w(\r)$
less than $\alpha$ where $\alpha$ is a small positive constant.
In addition to the above, we add the renormalization
of $\v(\r)$ and $\w(\r)$ at each step.
The above procedure does not guarantee the conservation of energy.
Although the increasing of energy at each step is small,
it may cause some influence for long time behavior.
For safety, we renormalize $\v(\r)$ and $\w(\r)$ just after the
vortex dynamics given by (~\ref{dyna}~).

Next we introduce the viscosity.
The viscosity is introduced as a diffusion process of the vorticity.
And  it is easily represented as;

\begin{equation}
\omega_z(\r_0,t+\Delta t)=(1-\nu\Delta t)\omega_z(\r_0,t)
+\sum_{\i=\ux,\uy,\uz} \omega_z(\r_0\pm\i,t)\nu\Delta t / 6 \label{visco}
\end{equation}
where $\nu$ is the viscosity. $y$ and $z$ components follow similar equations.
This is the conventional and the simplest discretization
of the diffusion equation.

In addition to the above, if we would like to maintain steady state,
we should introduce an external force.
The time development of $\w(\r)$ due to the force is very simple,
\begin{equation}
\omega_z(\r_0,t+\Delta t)=\omega_z(\r_0,t)+\rot\f(\r_0)\Delta t \label{force}
\end{equation}
where $\f$ is the  external force.

The whole process of our algorithm is
summarized as follows;

\begin{itemize}
\begin{enumerate}
\item Calculating $\sum_\r \v(\r)^2/2$
\item Deciding $\Delta t$
\item Time development corresponding to (~\ref{dyna}~)
\item Renormalizing $\v(\r),\w(\r)$ so as to keep
$\sum_\r \v(\r)^2/2$ constant.
\item Time development corresponding to the diffusion process.
(See (~\ref{visco}~))
\item Time development corresponding to the external force.
(See (~\ref{force}~))
\item Return to process 1.
\end{enumerate}
\end{itemize}

\section{The numerical results}
\label{results : sec}
In this section, we show some numerical results.
In the numerical simulation, we use the lattice with $L=24$ and
set $\alpha$=0.1.
External force $\f(\r)$ is given as
\def\sxp{\sin(k(x+\half))}
\def\syp{\sin(k(y+\half))}
\def\szp{\sin(k(z+\half))}
\def\cxp{\cos(k(x+\half))}
\def\cyp{\cos(k(y+\half))}
\def\czp{\cos(k(z+\half))}
\def\sx{\sin(kx)}
\def\sy{\sin(ky)}
\def\sz{\sin(kz)}
\def\cx{\cos(kx)}
\def\cy{\cos(ky)}
\def\cz{\cos(kz)}
\begin{eqnarray*}
f_x(\r)&=&4 \sxp\cy\cz\\
         &+&2 (\cxp\cy+\cy\cz+\sxp\sz)\\
f_y(\r)&=&-4\cx\syp\cz\\
         &+&2(\syp\sx+\cz\cyp+\cz\cx)\\
f_z(\r)&=&             2(\czp\cx+\cx\cy+\sy\szp)
\end{eqnarray*}
where $k = 2\pi/L$.
Actually, these correspond to the Taylor-Green vortex accompanied with the
components of wave number $2 \sqrt{2} \pi /L$.
This choice is not unique but, this is the case which contains the
fewest number of
kinds of wave number among those which recover
isotropy in the numerical simulations.

Remaining parameter is the viscosity $\nu$.
We investigate four values of $\nu$: $5,8,10,20 \times 10^{-3}$.
If we define microscopic Reynolds number
$R_\lambda = \sqrt{10/3}\langle\v^2/2\rangle/\nu\sqrt{\langle\w^2/2\rangle}$,
$R_\lambda\cong 5 \sim 20$
($\langle \cdots \rangle$ means average over the whole lattice point and time).
However we are not sure whether absolute values have meanings or not,
because in our model the unit cubic has already been coarse grained.

First we show the time development of the total $\sum_{\r} \v(\r)^2$ and
$\sum_{\r} \w(\r)^2$ in Fig \ref{times}. These behaviors look like those
of fully developed turbulece gotten with
the conventional direct integration (For example,
Ref. \cite{Sanada}).
Every results shown in below are averaged
over nine snap shots apart from each other by the
same intervals after the state reaches steady state
(Actually speaking, $600 < t < 3000$).
One can see total square velocities are
not influenced by the change of $\nu$.
On the other hand, the total square vorticities
are influenced by it strongly.

Before going forward, we show snapshot configurations
at a instance.
Fig. \ref{velocity} shows a velocity field snapshot.
Nothing abnormal is observed in this figure.
In Fig. \ref{vortex}, the bonds on which strong vorticities
exist are plotted. The number of plotted bonds are 527 which
is much less than total number of bonds : 41472.
In spite of small ratio of number of plotted bonds,
we can see large number of connected bonds.
This means our velocity field surely succeeds in presenting
coherent structure of vorticities which is observed in
direct integration \cite{Hosokawa88}.

For each calculations with four different
$\nu$, we need less than nine hours on CRAY-XMP (3000 steps each).
This is surprisingly short, if obtained velocity fields
really correspond to fully developed turbulence.

To check this point, three dimensional energy spectrums
are shown in Fig.\ref{spec}.
One must take care of that drastic decreasing
of energy spectrums on high wave number region is
due to not only the energy dissipation but also
the decreasing of number of mode having wave number of $k$.
So we should consider only $\mid k \mid \leq \pi$.
In order to see whether they exhibit inertial range,
that is, $k^{-5/3}$ law, the averaged spectrum is shown in
Fig. \ref{spec.av}.
For $\nu=10$ and $20 \times 10^{-3}$,
we can see inertial ranges.
However, for other two with smaller viscosity,
humps are seen in higher wave number region.
This means, for later two cases,
dynamics in small scale are damaged and
they cannot be regarded as a good
representation as highly developed
turbulence.

One may think these spectrum do not have any
dissipation ranges,
because decreasing of energy spectrum may be caused
by only insufficient number of modes as
pointed out in above.
For these four cases, we also calculate one dimensional energy spectrum
$E_z = < \mid \int v_z(x,y,z)exp(-ikx)dx\mid^2>_{y,z}$,
where $< \cdots >_{y,z}$ stands for the average over $y,z$.
We can see some scaling region obeying Kolmogorov's scaling region and
short inertial subrange.
Here we can see clear dissipation range. This time decreasing of
energy spectrum at the higher wave number is surely caused by the
energy dissipation because shortness of number of modes does not
occur in one dimension.

Therefore, we think
we can think cases with two larger viscosities
can be investigated as representing fully developed turbulence,
but not for other two cases.
However, we show results for all four cases for comparison.

In Figs.\ref{vx}, the probability distribution functions (PDF)
 of $\vx$ are shown.
They look like almost Gaussian which is also observed
in experiments and numerical calculation \cite{Kida89}.
In order to see the dependence upon $\nu$,
semi-logarithmic plot of them is shown in Fig.\ref{vx.log}.
Those having smaller $\nu$  are closer to Gaussian than others with
larger $\nu$.
This means even if higher wave number components are damaged,
lower component can properly behave and become closer to
fully developed turbulence.
This also reveals that our model works well to
produce developed turbulence.

Next several PDFs which are close to exponential-like distributions
are shown.
Figs.\ref{wx} to \ref{vx.high} show
PDF of $\wx$, $\vxx$,$\vxy$ and $\vx$ having only higher
wave number components $k \geq 2 \pi /3$.
All of them are normalized so as to have standard deviations of unity.
Roughly speaking all of them look like exponential distribution.
They agree with the results obtained before
\cite{Kida89,Metais,Hosokawa88,Hosokawa89,She,Gagne,VanAtta,Wyngaard} well.
Especially, asymmetry of PDF of $\vxx$ is properly
reproduced.
It is extremely surprising because these PDFs reflect the
property in the high wave number region which is
damaged for lower viscosities.
The fact means that our model can catch the essential feature
of turbulence which is hardly destroyed even by the
insufficient resolution.
It enables us to get some aspects on what the {\it essence} of
turbulence is~\cite{prl}.

These results clearly demonstrate our model works so well.
It is clear that our model can work very well to
represent fully developed turbulence although
it requires small memory-sizes and short cpu-time.

\section{Conclusion and Summary}
\label{disc : sec}

{}From the above results,
our model works remarkably well.
For example,
since our model has succeeded in reproducing some characteristic features
of fully developed turbulence; the Kolmogorov's scaling of the energy spectrum,
the development of energy and enstrophy, coherent structure of
vortex tubes,
 the qualitative feature of the distribution functions of the velocity and
 its differential we can conclude that we have achieved our purpose;
 to construct the simple model which can reproduce some qualitative
 features of turbulence.
We think it proves that the lattice model is also
valid in investigating the turbulence.
In contrast to the conventional direct integration
of Navier-Stokes equation,
our modeling is very tough to reproduce
the qualitative features of turbulence.
Our model is not damaged even with the
crude procedures like artificial rescaling in order to
conserve energy and rough discretization of the field,
which cause severe catastrophe to the direct integration.

Moreover we can expect many things for our model.
Since our model is so simple,
we can find some essential reason about the appearance of the
non-Gaussian PDFs,
which will be reported elsewhere\cite{prl}.
And even some rigorous calculation like estimating upper bounds of some value
may be possible, because many such calculation have been done on the lattice
 systems.
In addition to this, our model is connected with other lattice
models in statistical physics.
This will allow us to investigate several phenomena
like chemical reaction and phase ordering process under the existence
of turbulent flow.

\section{Acknowledgment}

Y-h. T. would like to thank Prof. H. M$\ddot{\rm u}$ller-Krumbhaar and Dr. W.
Zimmermann for helpful discussions and all the members of Theorie III at the
IFF for
their hospitalities. M. Takayasu is acknowledged by H.T. for useful
discussions.

\appendix{The cg method}
\label{ap1 : sec}
In this appendix, we give a brief explanation of the cg (conjugate gradient)
 method. For details, see references\cite{Hestenes,Togawa}.

Assume we have a set of linear equations.
In matrix form, $A\x=\b$, where $A$ is the coefficient matrix,
$\x$ is a vector consisting of variables and $\b$ is the constant vector
 having the same length as that of $\x$.
 The following procedures can be applied to the case that $A$ is not symmetric
 neither positive definite.

 The cg method is one of those which are suitable to search extreme values.
 If one would like to search the extreme values of the function $f(\r)$,
 one can follow along the flow $\nabla f(\r)$ iteratively.
 These are applicable to find the solution $A\x=\b$, if we define $f(\r)$ as
 $\r^2$ with $\r=A\x-\b$.
 To accelerate this process, in each iteration of the cg method,
 the direction of flow is tuned so as to be orthogonal to
 the flow directions in all the previous iterations.
 Therefore the cg method should converge within the $n$
 iterations, where $n$ is the number of components; that is,
 the total number of dimension in the variable space.
 Skipping detailed points, the actual algorithm of cg method is
\begin{itemize}
\begin{enumerate}
\item For given $k$-th approximate solution $\x_k$ and $k$-th
residue $\r_k$,
prepare a store array $\p_k$.
For $k$=1, $\p_1=\r_1=\b-A\x_1$
\item Calculate $\alpha_k=\frac{(A\p_k,\r_k)}{(A\p_k,A\p_k)}
=\frac{(A^T\r_k,A^T\r_k)}{(A\p_k,A\p_k)}$
\item Calculate the $k+1$-th approximate solution $\x_{k+1}=\x_k+\alpha_k\p_k$
\item Calculate the $k+1$-th residue
$\r_{k+1}=\b-A\x_{k+1}=\r_k-\alpha_k A \p_k$
\item Calculate $\beta_k = -\frac{(A\p_k,AA^T\r_{k+1})}{(A\p_k,A\p_k)}
=\frac{(A^T\r_{k+1},A^T\r_{k+1})}{(A^T\r_k,A^T\r_k)}$
\item Calculate $\p_{k+1}=A^T\r_{k+1}+\beta_{k}\p_k$
\end{enumerate}
\end{itemize}
For the convenience of numerical simulation,
one can do as followings using
store vectors $\g,\q$.
For $k=1$, $\r_1=\b-A\x_1,\p_0=0, c_0=1$, and
iterate like

$\g_k=A^T\r_k$

$c_k=(\g_k,\g_k)$

$\beta_k=c_k/c_{k-1}$

$\p_k=\g_k+\beta_k\p_{k-1}$

$\q_k=A\p_k$

$d_k=(\q_k,\q_k)$

$\alpha_k=c_k/d_k$

$\x_{k+1}=\x_k+\alpha_k\p_k$

$\r_{k+1}=\r_k-\alpha_k\q_k$

In our case, $\x_k$ consists of each component of velocity
$v_i(\r)$. Therefore if the linear dimension of the lattice is $L$,
the total number of the component $n$ is $3\times L^3$.
$A$ is coefficient matrix of (~\ref{rotv}~) and (~\ref{divv}~).
$\b$ is the vector containing all the components of $\w(\r)$.

\appendix{Relation with the vorticity equation}
\label{ap2 : sec}

One may think that our model lacks mathematical justification and
therefore cannot be trust.
In principle, we do not think our model needs any justifications,
because anyway our model is purposed only to reproduce qualitative results
at the moment.
If one has such a purpose, one dose not have to make his model being related
with
 the real system exactly.
This attitude is justified with the hope of existence of 'universality class'.
The universality class is a set of model or parameters which
can give the same results on some property.
One of the famous universality class in statistical physics is
'Ising universality class' in the magnetic phase transition.
Ising model is a set of magnetic moments which takes only values of $\pm1$,
therefore far from the true magnetic materials.
However, surprisingly, it can reproduce almost all
basic nature of 'Critical Phenomena'.
Although no one cannot prove a true magnetic material is exactly same
as the Ising model, one can get many deep insights about the
'true' phase transition accompanied with the critical phenomena.
And it has helped many people who would like to
know what the critical phenomena are.

We stand at the same point on the turbulence.
Therefore if our model can reproduce some qualitative
feature of turbulence, it is useful enough to use in order to
investigate turbulence.
In this meaning, we do not need any special justifications other than the
results itself.
However, we can actually shown some relationship with the true fluid equation
 as follows.

In order to expand (~\ref{dyna}~) in space,
first we expand $\v(\r)$ and $\w(\r)$ itself.
\begin{eqnarray*}
F_i(\r_0+\delta\ux+\delta\uy+\delta\uz)& =& \sum_{n=0}^{\infty}\frac{1}{n!}
(\dx\px+\dy\py+\dz\pz)^n F_i \\
                     & = & \exp(\dx\px+\dy\py+\dz\pz) F_i\\
                     &&
\end{eqnarray*}
where $F_i$ is either $v_i$ or $\omega_i$.
Considering the definition of $\tilde{v}, $we get from eq.(3),
\begin{eqnarray*}
\omega_i(t+\Delta t)&=& \omega_i(t)\\
&+&\Delta t[
 \cosh(\frac{\dj}{2} \tilde{\pj})\sinh(\dk\pk)
(-v_k \omega_i + 2\cosh(\frac{\di}{2}\pii ) v_i\omega_k)\\
& + & \cosh(\frac{\dk}{2} \tilde{\pk})\sinh(\dj\pj)
(-v_j \omega_i + 2\cosh(\frac{\di}{2}\pii ) v_i\omega_j)]
\end{eqnarray*}
where $i,j,k$ is $x,y,z$ respectively and change order cyclical
and the partial differentiation with tilde is applied to only the velocity
not to the vorticity.
And taking into account to the second order of $\di,\dj,\dk$ and the first
order of
$\delta t$, we can get
\begin{displaymath}
\frac{\partial\omega_i(t)}{\partial t}=
\dj\pj(v_i \omega_j - v_j \omega_i)
+\dk\pk(v_i \omega_k - v_k \omega_i)
\end{displaymath}
Using $\div \v =\div \w =0$ and setting $\dx=\dy=\dz=1$, this reduces to
\begin{displaymath}
\frac{\partial \w}{\partial t} + (\v \cdot \nabla ) \v = (\w \cdot \nabla)\v.
\end{displaymath}
This is the vorticity equation without viscosity term itself.
And it does not have 'any second order' terms.
This means we can introduce viscosity term
using eq.(~\ref{visco}~) correctly.
Therefore our model reduces to the dynamics of vortex
by taking the continuum limit.

Here we should mention our dynamics differ from the conventional ones.
In the conventional concepts, vortex tube should move with fluid in
nonviscostic case.
In our dynamics, as show in
Fig.~\ref{dyna} the center of vortex tube does not move.
In stead of that, the ${\em width}$\cite{width}
of the vortex tube will
increase rapidly.
We see that ours are better than conventional ones.
This is because the conventional ones give the second order term which
vanish in our dynamics.
The second order terms conflict with the true viscosity term
which is also  of the second order.
We guess this is the reason why the conventional vortex method
cannot either get good results nor consider viscosity correctly.
Therefore if one follows our dynamics,
vortex method can give us physical results.

\figure{ The relation of a vortex tube (the doubled vector) and
velocity vectors (the solid vectors).
$\omega$ is equal to $v_{1y}-v_{2y}-v_{2z}+v_{1z}$.
\label{lattice}}

\figure{ The actual changes of the distribution of "vortex tubes"
on the $\w$-lattice
 during the dynamical process.
If the part of the vortex tube has the vorticity $\omega$ and the velocities
on that part are $\tilde{v}_x,\tilde{v}_y$ (left), after the dynamics it has
branches of
 vortex tubes shown as narrow arrows(right).
The direction of the arrows shows the sign of the value of the vorticity.
And absolute values of the branches have two values depending on
whether it was induced by $\tilde{v}_x$ or $\tilde{v}_y$.
One must notice the vorticities on the part which has existed before dynamics
does not change at all.
This shows in our dynamics the canter of the vortex tube does not move and
 the vortex tube itself collapses automatically.
 In Appendix.~\ref{ap2 : sec} we can see this is better than
 the conventional picture.
\label{dyna : fig}}

\figure{Time development of $\v(\r)^2$(thin lines) and
$\w(\r)^2$(thick lines) (a) $\nu=5 \times 10^{-3}$
(b) $\nu=8 \times 10^{-3}$
(c) $\nu=10 \times 10^{-3}$
(d) $\nu=20 \times 10^{-3}$ \label{times}}

\figure{A velocity snap shot taken from $\nu=8 \times 10^{-3}$ at
$t \sim 1500$. Two dimensional cross section at $z=4$.
\label{velocity}}

\figure{Coherent structure of vortex tubes at
the same instance as that in Fig.~\protect\ref{velocity}.
Bonds with strong vorticities are plotted.
Summation over them is 14 percent of total square vorticities.
\label{vortex}}

\figure{Three dimensional energy spectrums.
(a) to (d) correspond to those in Fig. \ref{times}.
Solid lines indicate $k^{-5/3}$. \label{spec}}

\figure{Averaged spectrum corresponding to Figs. \ref{spec}. Whole range
is divided into 50 parts and spectrum is averaged within
each band.
\label{spec.av}}

\figure{The scaled energy spectrum.
$k_0=(\varepsilon/\nu^3)^{1/4}\pi/12$
 is proportional to
Kolmogorov wave number.
$\varepsilon=\nu \bigl< \vec\omega^{2}/2 \bigr >$ is dissipation energy.
Solid line indicates 5/3 law.
The symbols represent the cases with different values of
the viscosity rate $\nu$; $\times$ : $5 \times 10^{-3}$,
                     $\Box$ : $8 \times 10^{-3}$,
                     $\Diamond$ : $10 \times 10^{-3}$,
                     $+$ : $20 \times 10^{-3}$.
                     Error bars are smaller than the size of character.
\label{spec.1d}}

\figure{PDF of $\vx$. (a) to (d) correspond to those of Fig. \ref{times}.
The lines without chracter represent Gaussian distribution.
\label{vx}}

\figure{Semi-logarithmic plot of Fig. \ref{vx}.
$\times,\Box,\Diamond$ and $+$ correspond to (a),(b),(c) and (d), respectively.
\label{vx.log}}

\figure{PDF of $\wx$. Characters are the same as those of Fig.\ref{vx.log}
\label{wx}}

\figure{PDF of $\vxx$. Characters are the same as those of Fig.\ref{vx.log}
\label{vxx}}

\figure{PDF of $\vxy$. Characters are the same as those of Fig.\ref{vx.log}
\label{vxy}}

\figure{PDF of $\vx$.
Their lower wave number components
$k \leq 2 \pi /3$ are removed.
Characters are the same as those of Fig.\ref{vx.log}
\label{vx.high}}


\begin{references}
\bibitem{width}For example, the width of the vortex tube on $j-k$ surface can
be
defined as
$
W_i = \frac{\int dj dk \omega_{i}^{2}(j,k)\sqrt{(j-j_{0})^{2}+(k-k_{0})^{2}}}
{\int djdk \omega_i^{2}(j,k)}\nonumber
$
where $i,j$ and $k$ is $x,y$ and $z$ and changes the order cyclically.
$j_{0}$ and $k_{0}$ is the center of the vortex tube,
$
j_{0}=\frac{\int djdk \omega_{i}^{2}(j,k)j}
{\int dkdj \omega_{i}^{2}(j,k)}\nonumber.
$
\end{references}
\end{document}